# Effect of door delay on aircraft evacuation time

Martyn Amos and Andrew Wood

*Department of Computer Science, University of Exeter, Harrison Building, North Park Road, Exeter EX4 4QF, UK*

**The recent commercial launch of twin-deck Very Large Transport Aircraft (VLTA) such as the Airbus A380 has raised questions concerning the speed at which they may be evacuated [1]. The abnormal height of emergency exits on the upper deck has led to speculation that emotional factors such as fear may lead to door delay, and thus play a significant role in increasing overall evacuation time [2]. Full-scale evacuation tests are financially expensive and potentially hazardous, and systematic studies of the evacuation of VLTA are rare [3]. Here we present a computationally cheap agent-based framework [4] for the general simulation of aircraft evacuation [5-7], and apply it to the particular case of the Airbus A380. In particular, we investigate the effect of door delay, and conclude that even a moderate average delay can lead to evacuation times that exceed the maximum for safety certification. The model suggests practical ways to minimise evacuation time, as well as providing a general framework for the simulation of evacuation.**

The safe and rapid evacuation of passengers from aircraft during emergencies is clearly of paramount importance. In order for an aircraft to meet national and international regulations [8,9], and thus receive certification of its worthiness to fly, tests must be carried out to ensure that certain evacuation conditions are met. The most crucial of these conditions is that an aircraft must be capable of being fully evacuated of passengers and crew within 90 seconds, with only half of the emergency exits available. Full-scale live evacuation trials using volunteers are expensive to implement and potentially dangerous for the participants [3]. An aircraft's certification is therefore often based on the results of a single trial, and these results are usually kept confidential for commercial reasons. At the time of writing, no full evacuation test has been carried out for the Airbus A380. This aircraft is unusual in that it has a "double-decker" design, with two passenger decks stacked one on top of the other. This configuration raises particular issues concerning evacuation; specifically, the height of the upper exit doors from the ground. Fears have been raised that the abnormal height (7.9m) of the upper exits may lead to some passengers delaying their exit from the aircraft [2]. Door hesitation may prove costly in terms of the overall evacuation time, but we are unaware of any detailed and realistic studies examining the impact of this particular factor.

In order to augment live trials, several computer models have been developed to simulate the effect on overall evacuation speed of varying factors such as seat configuration [10], exit spacing [11] and passenger/crew behaviour [6], as well as events such as fire [7]. We propose, along similar lines, a model that may be used to quickly investigate the particular effect of exit door hesitation on aircraft evacuation time. We use an agent-based passenger modelling approach, combined with a grid-based environmental representation. The model was implemented in the NetLogo programming language [12], a standard multi-agent modelling environment. The internal structure of the aircraft is represented as a two-dimensional grid of "patches", each of which represents a real area of $0.25m^2$ (i.e., each patch is 0.5m by 0.5m). The front of the aircraft is to the right, with the port side at the top, and the starboard side at the bottom. The mechanism to implement realistic movement of passengers towards the exits is inspired by a model of slime-mold [13], which uses simple pheromone-based cues to facilitate cellular aggregation. Each patch is colour-coded according to its proximity to an exit (coloured red, two patches wide). The brightness of a patch is directly proportional to its proximity to an exit, and by defining the intensity of various patches we may highlight one or more routes from any square to the nearest exit. In this way, we may embed within the simulation signals corresponding to both the aircraft's floor-situated guidance lights and directional instructions issued by the flight crew. In order to conform to certification regulations, we choose three exits at random (the front and rear port exits, and the central starboard exit) and block them off with dark patches. Black patches represent either the gap between seat rows or the wall of the aircraft. An example of the initial configuration, corresponding to the upper deck of the Airbus A380, is depicted in Figure 1. We model only the upper deck, and assume no passenger movement between decks (this is consistent with the planned evacuation test). Passengers are represented as 199 individual entities (the small arrows in Figure 1) that are initially placed in seat squares, and which move around the grid in a probabilistic fashion. The figure of 199 represents full capacity for the A380 upper deck. Each passenger, $P_i$, has the following attributes defined: $p_i$ (current grid position), $h_i$ (heading), $m_i$ (maximum speed), $s_i$ (current speed), $d_i$ (door delay). The following global system variables are defined: $D$ (mean passenger door delay, measured in seconds), and $O$ (exit opening time). The value for $O$ represents the time taken for the cabin crew to open the exit doors and for emergency slides to be fully deployed, and is fixed at 14 seconds, based on previous studies [3]. The origin of the coordinate system used is at the centre of the aircraft, so internal grid positions run from -33 to 32 in the x-dimension, and from 5 to -5 in the y-dimension. Headings range from 0 to 360 (degrees), with the default being 90 (i.e., pointing towards the front of the aircraft). Door delay, $d_i$, is the time taken by an agent at an exit before leaving the aircraft, and is initialised as a random Poisson-distributed real number with mean $D$. In order to represent a full and realistic range of passenger ages and abilities, each agent is assigned a minimum speed of 0 and a randomly-generated, Poisson-distributed maximum speed between 0.3 m/s and 1.05 m/s, in line with previous studies [3].

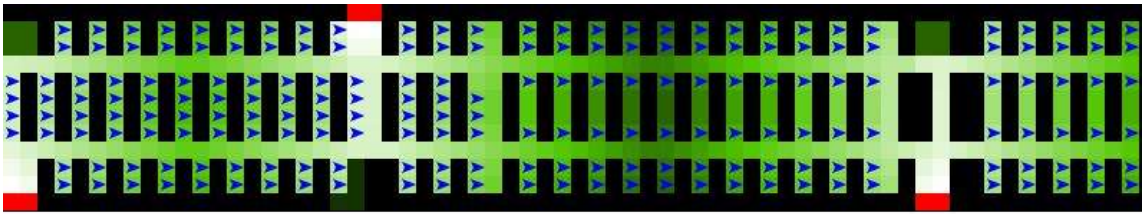

A time variable, $T$, is set to 0, and incremented by 1 every 0.1s. During each system "tick" the following occurs: first, the exits are checked. If an agent $A_i$ is on an exit (and $T*10 > O$), then $d_i$ is decremented by 0.1; if $d_i$ reaches 0 (and $T*10 >= O$) then the agent is considered to have exited the aircraft, and is removed from the simulation. Each agent $P_i$ is then instructed to move towards the nearest exit. This is achieved by $P_i$ adjusting its heading to try to move up the colour gradient defining the path to the nearest exit (that is, if a patch adjacent to an agent has a higher colour intensity than the one currently occupied, the agent will try to move to that patch). Adjacency is defined by the Moore neighbourhood [14] (i.e., the 8 surrounding patches). All movement is subject to the restriction that only one agent may occupy a given patch during a single tick. If the patch ahead is occupied by another agent $P_j$, then $P_i$ matches the speed of $P_j$ and decelerates (ensuring that $s_i$ does not fall below 0); if the patch is empty, then $P_i$ accelerates (ensuring that $s_i$ does not exceed $m_i$) and then moves forward a distance defined by $s_i$. This mechanism prevents agents from overtaking one another in the corridor, leading to gradual group deceleration and the formation of "traffic jams", as well as allowing agents to accelerate into available empty space.

The model was first calibrated with $D=0$. As data on actual evacuation tests are commercially-sensitive and not freely available, we used the results obtained by a previous study [3], which simulated a 90-second certification trial in an aircraft of similar layout carrying 236 passengers on the upper deck. The average evacuation time in these simulated trials was 64.1s (range 59.2 – 72.7). With a 16% smaller passenger capacity, we would expect to obtain an average evacuation time in the vicinity of (0.84 * 64.1 = 54s). The average evacuation time for our model, over 100 trials, was 57.9s (std. dev. 1.50), suggesting that the base model is valid. We observed realistic traffic flow patterns, with passengers making full use of alternative routes to the exit, where available (movies illustrating this are supplied with the supplementary materials.)

We then investigated the effect of increasing average door delay, D. We carried out 100 trials for each value $0.1 \leq D \leq 1.5$ (in intervals of 0.1). Our findings are plotted in Figure 2. Based on these results, we predict that the 90s certification threshold for the Airbus A380 will be exceeded if average door delay, *D*, exceeds 1.1s. A small-scale study [2], using a double-deck mock-up with 42 passengers and one exit, yielded an average door delay of 1.53s. Increasing *D* to 1.5s within our model gives an average evacuation time of 106.40s (118% of the certification threshold time.)

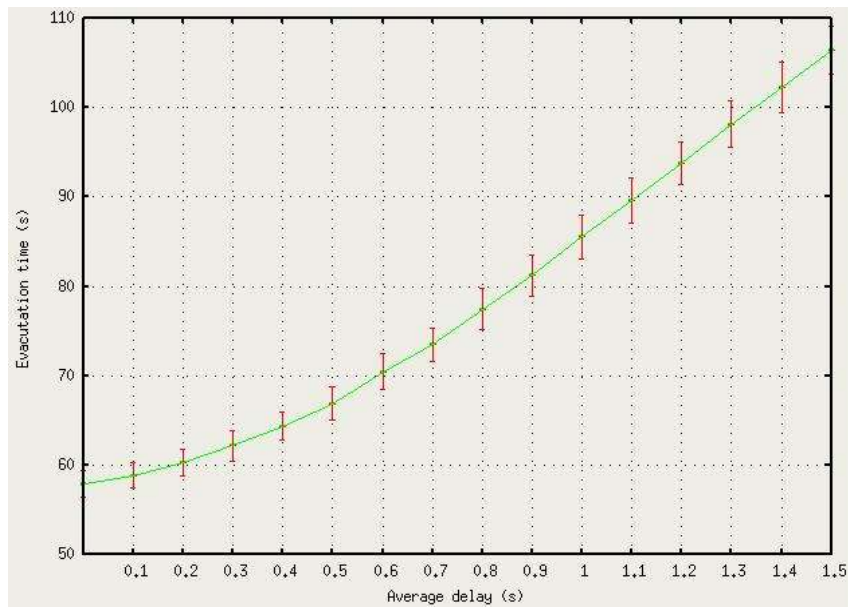

We now comment on generic aspects of our simulation framework. The model is computationally cheap to run; in the worst case (i.e., where $D = 1.5$), 100 trials were completed in 395 seconds (under 4 seconds per trial) on an NEC Versa laptop (140 GHz Intel Pentium, 224 Mb of RAM) running Microsoft Windows XP. It is also straightforward to add extra environmental or passenger-specific factors to the model, with little additional computational cost (for example, we have included a "general

6uncertainty" variable in the model in order to study the effect of smoke or failure of cabin warning systems, but this was unused in the current study).

Our agent-based evacuation model is based on plausible assumptions, and generates realistic overall behaviour using a minimal set of variables. Having calibrated the model according to available data, we are able to reproduce the findings of previous studies of aircraft evacuation. We therefore conclude that it is suitable for drawing conclusions about the effect of exit door hesitation on the evacuation of aircraft. In addition, the use of environmental cues based on pheromones is a novel aspect of our simulation which may well find broader application. The main implication of our findings is that door delay must be minimised for effective evacuation, and this impacts on airline policy on cabin crew training and passenger briefing. We now call for complementary data from live evacuations in order to test our model quantitatively.

Authors declare they have no competing financial interests. A.W. contributed to model design and implementation. M.A. contributed to model design and implementation, carried out experiments and wrote the paper.

Correspondence and requests for materials should be addressed to M.A. (M.R.Amos@ex.ac.uk).

Figure 1: Initial configuration of simulation. The internal layout of the aircraft is defined by a grid of squares, with the port side at the top of the screen. Each square is coloured according to its proximity to an exit, with the brightest squares being closest. Passengers are represented by arrows. This example depicts the situation where half of the exits are unavailable (rear and front port side, middle starboard side).

Figure 2: Simulation results for different values of average door delay, *D*. Each data point shows the average evacuation time over 100 trials for a particular value of *D*. Error bars show standard deviation.